\documentclass[useAMS,usenatbib]{mn2e}

\usepackage{graphicx}
\usepackage{amssymb}

\begin{document}

\title[Dust particles in MMRs influenced by an IGF]
{Dust particles in mean motion resonances influenced by
an interstellar gas flow}
\author[P. P\'{a}stor]{
P.~P\'{a}stor\thanks{pavol.pastor@hvezdarenlevice.sk}\\
Tekov Observatory, Sokolovsk\'{a} 21, 934~01 Levice, Slovak Republic}

\date{}

\pagerange{\pageref{firstpage}--\pageref{lastpage}} \pubyear{2012}

\maketitle

\label{firstpage}

\begin{abstract}
The orbital evolution of a dust particle captured in a mean motion resonance
with a planet in circular orbit under the action of the Poynting-Robertson
effect, radial stellar wind and an interstellar gas flow of is investigated.
The secular time derivative of Tisserand parameter is analytically derived
for arbitrary orbit orientation. From the secular time derivative
of Tisserand parameter a general relation between the secular time
derivatives of eccentricity and inclination is obtained.
In the planar case (the case when the initial dust particle position vector,
initial dust particle velocity vector and interstellar gas velocity vector
lie in the planet orbital plane) is possible to calculate directly
the secular time derivative of eccentricity.

Using numerical integration of equation of motion we confirmed our
analytical results in the three-dimensional case and also in the planar case.
Evolutions of eccentricity of the dust particle captured in an exterior
mean motion resonance under the action of the Poynting-Robertson effect,
radial stellar wind for the cases with and without the interstellar gas
flow are compared.

Qualitative properties of the orbital evolution in the planar case are
determined. Two main groups of the secular orbital evolutions exist.
In the first group the eccentricity and argument of perihelion approach
to some values. In the second group the eccentricity oscillates and
argument of perihelion rapidly shifts.
\end{abstract}

\begin{keywords}
ISM: general -- celestial mechanics -- interplanetary medium
\end{keywords}

\section{Introduction}
\label{sec:intro}

For the analysis of dynamical evolution of dust particles
in the vicinity of a star is necessary to take into account also
non-gravitational effects. From the non-gravitational effects accelerations
caused by the electromagnetic radiation and the corpuscular
radiation (stellar wind) of the central star is most often considered.
Influence of the electromagnetic radiation on the dynamics
of dust grains is usually described using Poynting-Robertson (PR) effect
\citep{poynting,robertson,PRI,klacka2004,Icarus}. For the corpuscular
radiation is usually assumed radial propagation from the central star
\citep{comet}. A more general form of the acceleration caused by
the corpuscular radiation can be found in \citet{covsw}.
When orbital periods of the particle and a planet are in a ratio of small
natural numbers mean motion resonances (MMRs) can occur.
If the dust particle is captured in an MMR, changes of the semimajor
axis caused by small non-gravitational effects are balanced by the resonant
interaction with the planet's gravity field. \cite{shepherd}
predicted a ring of dust particles orbiting the Sun captured
in mean motion resonances with planet Earth. This ring was confirmed
by observations from satellites IRAS \citep{ring,IRAS} and COBE
\citep{COBE}. The paper \cite{shepherd} was followed by many others
who investigated behaviour of dust particles captured in MMRs during the last
three decades \citep*[e.g.][]{WJ,MV,SN,LZ1995,LZJ,LZ1997,
LZ1999,MM,holmes,KH,DM,nsp1,nsp2,krivov,SK,reach,MW,ertel}. If the planet
moves in a circular orbit around the Sun and mass of the particle
is negligible in comparison with mass of the Sun and also in comparison
with mass of the planet, we have a special gravitational problem of three
bodies. This gravitational problem is called the circular restricted
three-body problem (CR3BP) in celestial mechanics.
An analytic expression for the secular time derivative of orbital
eccentricity of the dust particle captured in an MMR in the planar
CR3BP with the PR effect and the radial solar wind was found in
\citet{LZ1997}. Due to the rotation of the star stellar wind can be
non-radial in general. According to Helios 2 measurements \citep{bruno}
the angle between the radial direction and the direction
of solar wind velocity is approximately constant (at least for
the distances covered by observations). The secular time derivative
of orbital eccentricity of the dust particle captured in an MMR
in the planar CR3BP with the PR effect and such non-radial solar
wind was calculated in \citet{AA2} and all possibilities of the secular
eccentricity evolution were analytically determined in \citet{AA3}.
The secular evolution of orbital eccentricity and argument of
perihelion of the dust particle captured in an MMR in the planar circular
and elliptical restricted three-body problem with the PR effect was
numerically investigated in \citet*{CMDA}.

Interstellar medium atoms penetrate into the Solar System due to
relative motion of the Solar System with respect to the interstellar medium.
These approaching atoms form an interstellar gas flow (IGF). Influence
of this IGF on motion of dust particles orbiting a star was
mentioned already in \citet{comet}. Motion of dust particles orbiting
the Sun influenced by the IGF was analytically and numerically
investigated in \citet{scherer}. When the IGF velocity
vector lies in the orbital plane of the particle and the particle is under
the action of the PR effect, radial solar wind, and an IGF,
then the motion occurs in a plane. \citet{scherer} has correctly
described qualitative properties of the shift of perihelion
in the planar case despite of the fact that his calculations
contain several errors. Secular time derivatives of
semimajor axis, eccentricity and argument of perihelion in the planar
case were calculated in \cite{NOVA}. In \citet*{flow} secular time
derivatives of all Keplerian orbital elements for arbitrary orientation
of the orbit with respect to interstellar gas velocity vector were
calculated. The secular time derivatives of orbital elements were in
\citet{flow} derived under the assumptions (a) that the acceleration caused
by the IGF is small compared to the gravitation of
a central object, (b) that the speed of the IGF is large
in comparison with the speed of the dust particle (speeds are
determined with respect to the central object) and (c) that
the speed of the IGF is large also in comparison with
the mean thermal speed of the gas in the flow (Mach number $\gg$ 1).
Under these assumptions the IGF always causes decrease
of the secular semimajor axis of the dust particle. This decrease of
the secular semimajor axis was confirmed analytically
in \citet{bera} and using numerical integrations in \citet{flow},
\citet{marthe} and \citet{marzari}. The acceleration
of the dust particle caused by the IGF depends
on the drag coefficient which is, for the given particle and the flow
of interstellar gas, a specific function of the relative speed of the dust
particle with respect to the interstellar gas \citep*{baines}.
The drag coefficient was in \citet{scherer}, \citet{NOVA}, \citet{flow},
\citet{marthe} and \citet{marzari} taken into account as a constant.
This is a consequence of assumption (c). The derivation of the secular
time derivatives of all Keplerian orbital elements was generalised by
adding variability of the drag coefficient during orbit in \citet{dyncd}
using a method applied in \citet{bera} for the secular time derivative
of semimajor axis. In view of assumptions this means that
the secular time derivatives were derived under the assumptions (a), (b)
and assumption (c) was replaced by assumption that the mean thermal
speed of the gas in the flow is not close to zero (see \citealt{baines}).
Also under these assumptions the secular semimajor axis of particle's orbit
always decreases under the action of the IGF. The minimal
and maximal values of the decrease of the semimajor axis were also
determined in \citet{dyncd}.

In this work we use the secular time derivatives of orbital elements
caused by an IGF derived in \citet{dyncd} to obtain some basic properties of
orbital evolution of the dust particle captured in an MMR with a planet
in circular orbit.

\section{Secular orbital evolution in MMRs}
\label{sec:MMR}

Within the framework of the CR3BP under the assumptions that
the particle is far enough from the planet and mass of the planet is
negligible with respect to the mass of the Sun F. F. Tisserand found
a quantity which remains constant during the motion of the particle
(\citealt{tisserand}, e.g. \citealt{brocle})
\begin{equation}\label{CT0}
C_{T0} = \frac{1}{2 a} + \sqrt{\frac{a ( 1 - e^{2} )}{a_{P}^{3}}} \cos i ~,
\end{equation}
where $a$ is the semimajor axis of the particle orbit,
$e$ the eccentricity of the particle orbit,
$i$ is the inclination of the particle orbit with respect
to the planet orbital plane and $a_{P}$ is the semimajor axis of the planet
orbit. If we take into account also non-gravitational effects,
Tisserand parameter will no longer be constant, in general.
From the non-gravitational effects the electromagnetic radiation of the Sun
in the form of the PR effect is most often considered. If we add Keplerian
term of the PR effect to the central Keplerian acceleration of the Sun,
Tisserand parameter obtains the following form (cf. \citealt{LZ1997})
\begin{equation}\label{CT}
C_{T} = \frac{1 - \beta}{2 a_{\beta}} +
\sqrt{\frac{( 1 - \beta ) a_{\beta} ( 1 - e_{\beta}^{2} )}{a_{P}^{3}}}
\cos i_{\beta} ~.
\end{equation}
The parameter $\beta$ is defined as the ratio of the electromagnetic
radiation pressure force and the gravitational force between the Sun and
the particle at rest with respect to the Sun
\begin{equation}\label{beta}
\beta = \frac{3 L_{\odot} \bar{Q}'_{pr}}{16 \pi c \mu R \varrho} ~.
\end{equation}
Here, $L_{\odot}$ is the solar luminosity, $\bar{Q}'_{pr}$ is
the dimensionless efficiency factor for radiation pressure averaged
over the solar spectrum and calculated for the radial direction
($\bar{Q}'_{pr}$ $=$ 1 for a perfectly absorbing sphere),
$c$ is the speed of light in vacuum, $\mu$ $=$ $G M_{\odot}$,
$G$ is the gravitational constant, $M_{\odot}$ is the mass of the Sun,
and $R$ is radius of the dust particle with the mass density $\varrho$.
Subscript $\beta$ in Eq. (\ref{CT}) denotes that osculating
orbital elements are calculated using acceleration
$- \mu ( 1 - \beta ) \vec{r} / r^{3}$ as the central
Keplerian acceleration. Here, $\vec{r}$ is the position vector
of the dust particle with respect to the Sun and $r$ $=$
$\vert \vec{r} \vert$.

The particle is in an MMR with the planet when the ratio of their
mean motions is equal to the ratio of two small natural numbers. For an
exterior $q$-th order MMR we have $n_P / n$ $=$ $( p + q ) / p$ and
$n_P / n$ $=$ $p / ( p + q )$ for an interior $q$-th order MMR.
Here, $p$ and $q$ are two small natural numbers, $n_P$ is the mean
motion of the planet and $n$ is the mean motion of the particle.
The special case mean motion 1/1 resonance corresponds to $q$ $=$ 0.
If the dust particle is captured in the MMR, then the semimajor axis
of the particle's orbit librates around a constant value. This can be
written in the following form
\begin{equation}\label{constant}
\left \langle \frac{da_{\beta}}{dt} \right \rangle = 0 ~.
\end{equation}
Averaging in Eq. (\ref{constant}) is over a period of the resonant libration.
Using Kepler's Third Law we can obtain a relation between the semimajor axis
of the planet and the semimajor axis of the particle captured in the MMR.
We have
\begin{eqnarray}
\label{KTL_planet}
G ( M_{\odot} + M_{P} ) &=& n_{P}^{2} a_{P}^{3} ~, \\
\label{KTL_particle}
G M_{\odot} ( 1 - \beta ) &=& n^{2} a_{\beta}^{3} ~,
\end{eqnarray}
where $M_{P}$ is the mass of the planet. Putting Eq. (\ref{KTL_planet})
into Eq. (\ref{KTL_particle}) (with assumption $M_{P}$ $\ll$ $M_{\odot}$)
we find the relation
\begin{equation}\label{axes}
a_{\beta} = a_{P} \left ( 1 - \beta \right )^{1/3}
\left ( \frac{n_{P}}{n} \right )^{2/3} ~.
\end{equation}
Now, we admit that the dust particle is under the action of arbitrary
non-gravitational effects for which secular time derivatives
of the Keplerian orbital elements can be determined. Because, the secular
semimajor axis is constant, the gravitational influence of the planet
(subscript $G$) must compensate changes of the semimajor
axis caused by the non-gravitational effects (subscript $EF$)
\begin{equation}\label{ressemimajor}
\left \langle \frac{da_{\beta}}{dt} \right \rangle =
\left \langle \frac{da_{\beta}}{dt} \right \rangle_{G} +
\left \langle \frac{da_{\beta}}{dt} \right \rangle_{EF} = 0 ~.
\end{equation}
We have also
\begin{equation}\label{reseccentricity}
\left \langle \frac{de_{\beta}}{dt} \right \rangle =
\left \langle \frac{de_{\beta}}{dt} \right \rangle_{G} +
\left \langle \frac{de_{\beta}}{dt} \right \rangle_{EF} ~,
\end{equation}
\begin{equation}\label{resinclination}
\left \langle \frac{di_{\beta}}{dt} \right \rangle =
\left \langle \frac{di_{\beta}}{dt} \right \rangle_{G} +
\left \langle \frac{di_{\beta}}{dt} \right \rangle_{EF} ~.
\end{equation}
For a particle with $\beta$ $=$ 0 (e.g., an asteroid) in the planar
CR3BP ($i_{\beta}$ $=$ 0 and $di_{\beta} / dt$ $=$ 0) captured in an
MMR the eccentricity oscillates around a constant value,
because the Tisserand parameter (Eq. \ref{CT0}) also oscillates around
a constant value. Now we find a relation between the secular evolution
of eccentricity and inclination for the dust particle under the action
of the non-gravitational effects captured in an MMR. The total differential
of the Tisserand parameter is
\begin{eqnarray}\label{total}
\left \langle \frac{dC_{T}}{dt} \right \rangle &=&
      \frac{\partial C_{T}}{\partial a_{\beta}}
      \left \langle \frac{da_{\beta}}{dt} \right \rangle +
      \frac{\partial C_{T}}{\partial e_{\beta}}
      \left \langle \frac{de_{\beta}}{dt} \right \rangle +
      \frac{\partial C_{T}}{\partial i_{\beta}}
      \left \langle \frac{di_{\beta}}{dt} \right \rangle
\nonumber \\
&=&   \frac{\partial C_{T}}{\partial a_{\beta}}
      \left ( \left \langle \frac{da_{\beta}}{dt} \right \rangle_{G} +
      \left \langle \frac{da_{\beta}}{dt} \right \rangle_{EF} \right )
\nonumber \\
& &   + ~\frac{\partial C_{T}}{\partial e_{\beta}}
      \left ( \left \langle \frac{de_{\beta}}{dt} \right \rangle_{G} +
      \left \langle \frac{de_{\beta}}{dt} \right \rangle_{EF} \right )
\nonumber \\
& &   + ~\frac{\partial C_{T}}{\partial i_{\beta}}
      \left ( \left \langle \frac{di_{\beta}}{dt} \right \rangle_{G} +
      \left \langle \frac{di_{\beta}}{dt} \right \rangle_{EF} \right ) ~.
\end{eqnarray}
When the particle is far enough from the planet and the mass of the planet is
negligible with respect to the mass of the Sun, then the planet gravitation
does not change value of the Tisserand parameter. Hence
\begin{eqnarray}\label{gravitation}
\left \langle \frac{dC_{T}}{dt} \right \rangle &=&
      \frac{\partial C_{T}}{\partial a_{\beta}}
      \left \langle \frac{da_{\beta}}{dt} \right \rangle_{EF} +
      \frac{\partial C_{T}}{\partial e_{\beta}}
      \left \langle \frac{de_{\beta}}{dt} \right \rangle_{EF}
\nonumber \\
& &   + ~\frac{\partial C_{T}}{\partial i_{\beta}}
      \left \langle \frac{di_{\beta}}{dt} \right \rangle_{EF} ~.
\end{eqnarray}
Using the condition that the particle is in an MMR (Eq. \ref{constant})
and Eq. (\ref{gravitation}) in Eq. (\ref{total}) we obtain
\begin{eqnarray}\label{full}
\left \langle \frac{dC_{T}}{dt} \right \rangle &=&
      \frac{\partial C_{T}}{\partial e_{\beta}}
      \left \langle \frac{de_{\beta}}{dt} \right \rangle +
      \frac{\partial C_{T}}{\partial i_{\beta}}
      \left \langle \frac{di_{\beta}}{dt} \right \rangle
\nonumber \\
&=&   \frac{\partial C_{T}}{\partial a_{\beta}}
      \left \langle \frac{da_{\beta}}{dt} \right \rangle_{EF} +
      \frac{\partial C_{T}}{\partial e_{\beta}}
      \left \langle \frac{de_{\beta}}{dt} \right \rangle_{EF}
\nonumber \\
& &   + ~\frac{\partial C_{T}}{\partial i_{\beta}}
      \left \langle \frac{di_{\beta}}{dt} \right \rangle_{EF} ~.
\end{eqnarray}
Eq. (\ref{full}) represents the relation which must hold between the secular
time derivatives of eccentricity and inclination of the dust particle
captured in the MMR under the action of the non-gravitational effects.
For the secular time derivative of eccentricity we obtain
\begin{eqnarray}\label{ratios}
\left \langle \frac{de_{\beta}}{dt} \right \rangle &=&
      \frac{\partial C_{T} / \partial a_{\beta}}
      {\partial C_{T} / \partial e_{\beta}}
      \left \langle \frac{da_{\beta}}{dt} \right \rangle_{EF} +
      \left \langle \frac{de_{\beta}}{dt} \right \rangle_{EF}
\nonumber \\
& &   - ~\frac{\partial C_{T} / \partial i_{\beta}}
      {\partial C_{T} / \partial e_{\beta}}
      \left \langle \frac{di_{\beta}}{dt} \right \rangle_{G} ~.
\end{eqnarray}
Calculation of the partial derivatives yields
\begin{eqnarray}\label{general}
\left \langle \frac{de_{\beta}}{dt} \right \rangle &=&
      \frac{1 - e_{\beta}^{2}}{2 a_{\beta} e_{\beta}}
      \left [ \frac{( 1 - \beta )^{1/2}}{( 1 - e_{\beta}^{2} )^{1/2}
      \cos i_{\beta} }
      \left ( \frac{a_{P}}{a_{\beta}} \right )^{3/2} - 1 \right ]
\nonumber \\
& &   \times ~\left \langle \frac{da_{\beta}}{dt} \right \rangle_{EF} +
      \left \langle \frac{de_{\beta}}{dt} \right \rangle_{EF}
\nonumber \\
& &   - ~\frac{1 - e_{\beta}^{2}}{e_{\beta}} \tan i_{\beta}
      \left \langle \frac{di_{\beta}}{dt} \right \rangle_{G} ~.
\end{eqnarray}
Due to the derivation formulated for arbitrary non-gravitational effects with
known secular time derivatives of orbital elements Eqs. (\ref{full}) and
(\ref{general}) are generalisations of the results obtained by \citet{LZ1997}.
\begin{figure*}
\begin{center}
\includegraphics[width=0.9\textwidth]{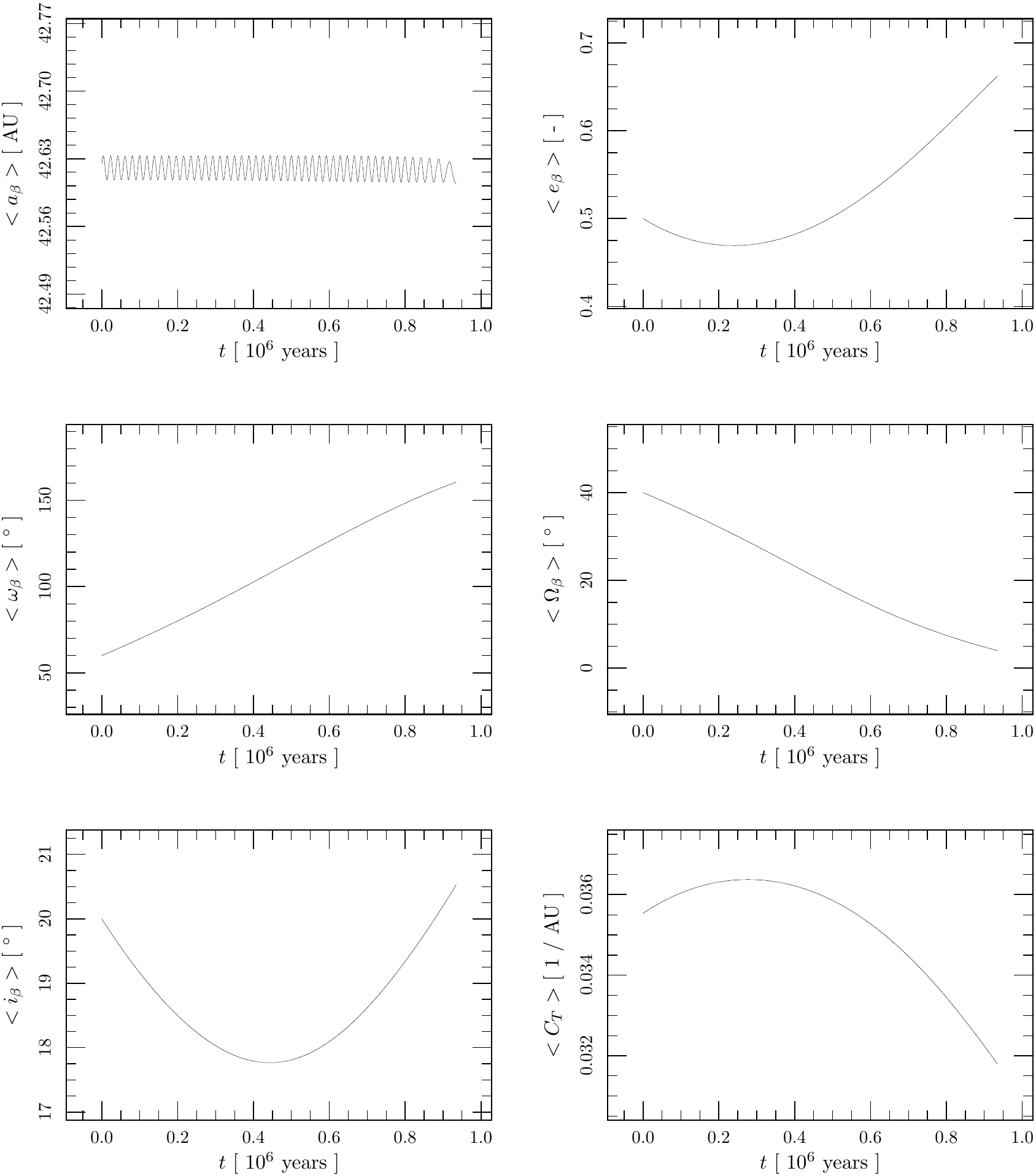}
\end{center}
\caption{Evolutions of the orbit averaged semi-major axis, eccentricity,
argument of perihelion, longitude of the ascending node, inclination
and Tisserand parameter of a dust particle with $R$ $=$ 2 $\mu$m,
$\varrho$ $=$ 1 g/cm$^{3}$, and $\bar{Q}'_{pr}$ $=$ 1
captured in an exterior mean motion orbital 2/1 resonance with
Neptune under the action of the PR effect, radial solar wind and IGF.}
\label{F1}
\end{figure*}
\begin{figure*}
\begin{center}
\includegraphics[width=0.9\textwidth]{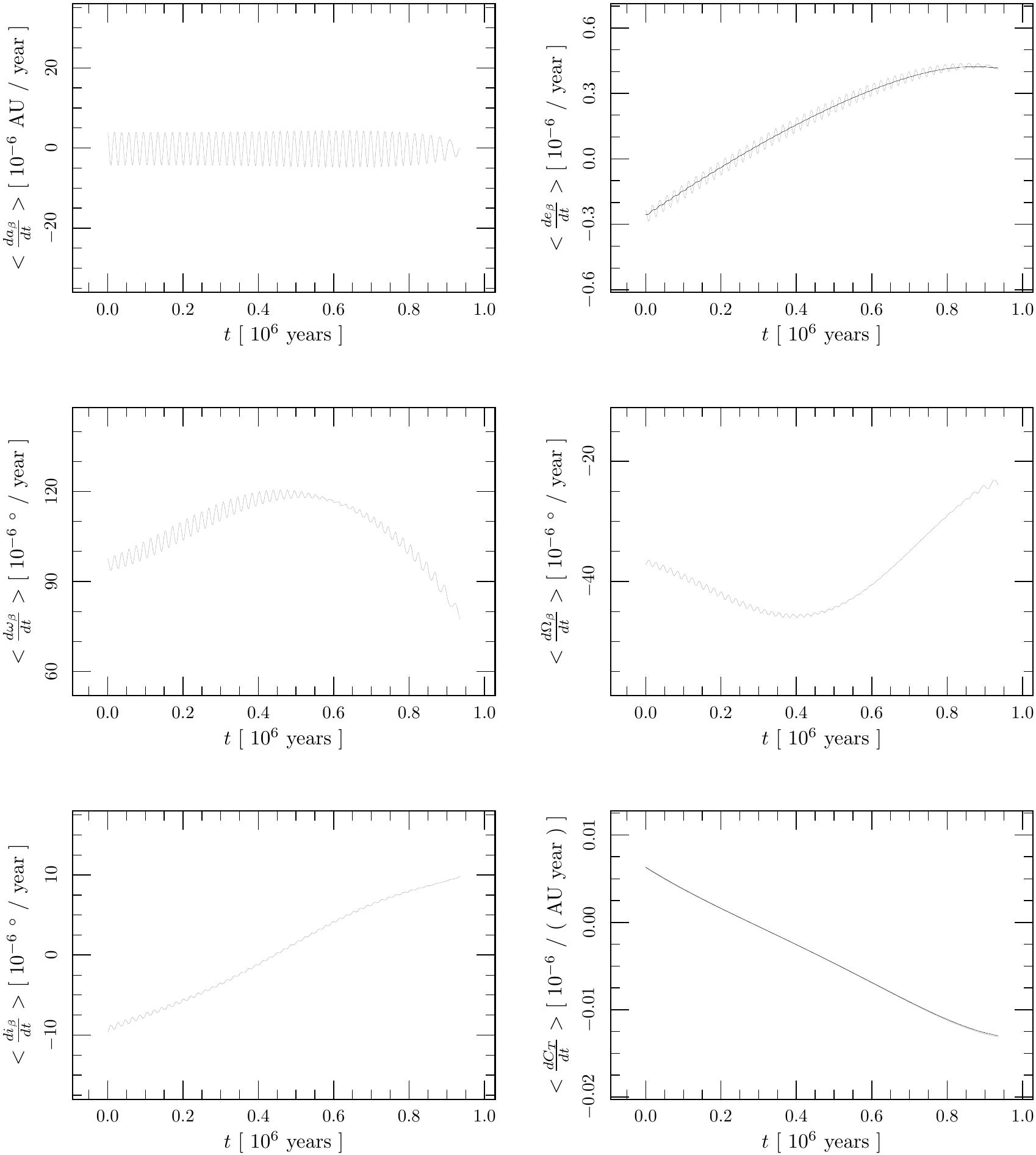}
\end{center}
\caption{Evolutions of the orbit averaged time derivative of semi-major axis,
eccentricity, argument of perihelion, longitude of the ascending node,
inclination and Tisserand parameter during the numerical solution of
Eq. (\ref{EOM}) depicted in Fig. \ref{F1} (grey line). The evolution
of the orbit averaged time derivative of Tisserand parameter is compared
with the secular time derivative of Tisserand parameter obtained
from Eq. (\ref{evolutions}) (black line). The evolution of the orbit averaged
time derivative of eccentricity is compared with the secular time
derivative of eccentricity calculated from Eq. (\ref{evolutions}) using
the numerical values of $\langle di_{\beta} / dt \rangle$ depicted
in the bottom left panel (black line).}
\label{F2}
\end{figure*}

\section{Influence of an IGF on orbital evolution of dust particles in MMRs}
\label{sec:secular}

In this section we use the analytical theory from the previous section in
order to find equations for the secular time derivatives of the particle's
orbital elements in the case when the particle, captured in an MMR, is under
the action of the PR effect, radial solar wind and IGF. The secular changes
caused by the IGF will be described using the secular time derivatives
of orbital elements derived in \citet{dyncd}. For the PR effect and radial
solar wind the standard expressions will be used \citep{WW,covsw}. The secular
time derivatives of the orbital elements caused by these effects then are
\begin{eqnarray}
\label{dadt_sys}
\left \langle \frac{da_{\beta}}{dt} \right \rangle_{EF} &=& - \beta
      \frac{\mu}{c} \left ( 1 + \frac{\eta}{\bar{Q}'_{pr}} \right )
      \frac{2 + 3e_{\beta}^{2}}{a_{\beta} ( 1 - e_{\beta}^{2} )^{3/2}}
\nonumber \\
& &   - ~\sum_{i} 2 a_{\beta} c_{0i} \gamma_{i} v_{F}^{2}
      \sqrt{\frac{p_{\beta}}{\mu \left ( 1 - \beta \right )}} \sigma_{\beta}
\nonumber \\
& &   \times ~\Biggl [ 1 + \frac{g_{i}}{v_{F}^{2}}
      \frac{1 - \sqrt{1 - e_{\beta}^{2}}}{e_{\beta}^{2}}
\nonumber \\
& &   \times ~\left ( S_{\beta}^{2} +
      I_{\beta}^{2} \sqrt{1 - e_{\beta}^{2}} \right ) \Biggr ] ~,\\
\label{dedt_sys}
\left \langle \frac{de_{\beta}}{dt} \right \rangle_{EF} &=& - \beta
      \frac{\mu}{c} \left ( 1 + \frac{\eta}{\bar{Q}'_{pr}} \right )
      \frac{5 e_{\beta}}{2 a_{\beta}^{2} ( 1 - e_{\beta}^{2} )^{1/2}}
\nonumber \\
& &   + ~\sum_{i} c_{0i} \gamma_{i}
      v_{F} \sqrt{\frac{p_{\beta}}{\mu \left ( 1 - \beta \right )}}
\nonumber \\
& &   \times ~\Biggl [ \frac{3I_{\beta}}{2} +
      \frac{\sigma_{\beta} g_{i} ( I_{\beta}^{2} - S_{\beta}^{2} )
      ( 1 - e_{\beta}^{2} )}{v_{F}e_{\beta}^{3}}
\nonumber \\
& &   \times ~\left ( 1 - \frac{e_{\beta}^{2}}{2} -
      \sqrt{1 - e_{\beta}^{2}} \right ) \Biggr ] ~,\\
\label{domegadt_sys}
\left \langle \frac{d \omega_{\beta}}{dt} \right \rangle_{EF} &=& \sum_{i}
      \frac{c_{0i} \gamma_{i} v_{F}}{2}
      \sqrt{\frac{p_{\beta}}{\mu \left ( 1 - \beta \right )}}
      \Biggl \{ - \frac{3S_{\beta}}{e_{\beta}}
\nonumber \\
& &   + ~\frac{\sigma_{\beta} g_{i} S_{\beta}I_{\beta}}{v_{F}e_{\beta}^{4}}
      \biggl [ e_{\beta}^{4} - 6e_{\beta}^{2} + 4 -
      4(1 - e_{\beta}^{2})^{3/2} \biggr ]
\nonumber \\
& &   + ~C_{\beta} \frac{\cos i_{\beta}}{\sin i_{\beta}}
      \biggl [ \frac{3e_{\beta} \sin \omega_{\beta}}{1 - e_{\beta}^{2}} -
      \frac{\sigma_{\beta} g_{i}}{v_{F}}
\nonumber \\
& &   \times ~( S_{\beta} \cos \omega_{\beta} -
      I_{\beta} \sin \omega_{\beta} ) \biggr ] \Biggr \} ~,\\
\label{dOmegadt_sys}
\left \langle \frac{d \Omega_{\beta}}{dt} \right \rangle_{EF} &=& \sum_{i}
      \frac{c_{0i} \gamma_{i} v_{F} C_{\beta}}{2 \sin i_{\beta}}
      \sqrt{\frac{p_{\beta}}{\mu \left ( 1 - \beta \right )}}
\nonumber \\
& &   \times ~\Biggl [ - \frac{3e_{\beta} \sin \omega_{\beta}}
      {1 - e_{\beta}^{2}} + \frac{\sigma_{\beta} g_{i}}{v_{F}}
\nonumber \\
& &   \times ~( S_{\beta} \cos \omega_{\beta} -
      I_{\beta} \sin \omega_{\beta} ) \Biggr ] ~,\\
\label{didt_sys}
\left \langle \frac{di_{\beta}}{dt} \right \rangle_{EF} &=& - \sum_{i}
      \frac{c_{0i} \gamma_{i} v_{F} C_{\beta}}{2}
      \sqrt{\frac{p_{\beta}}{\mu \left ( 1 - \beta \right )}}
\nonumber \\
& &   \times ~\Biggl [ \frac{3e_{\beta} \cos \omega_{\beta}}
      {1 - e_{\beta}^{2}} + \frac{\sigma_{\beta} g_{i}}{v_{F}}
\nonumber \\
& &   \times ~( S_{\beta} \sin \omega_{\beta} +
      I_{\beta} \cos \omega_{\beta} ) \Biggr ] ~,
\end{eqnarray}
where $\omega_{\beta}$ is the argument of perihelion
and $\Omega_{\beta}$ is the longitude of the ascending node.
The sums in Eqs. (\ref{dadt_sys})-(\ref{didt_sys}) run over all particle
species $i$ in the IGF. $\eta$ is the ratio of solar wind energy to
electromagnetic solar energy, both radiated per unit of time
\begin{equation}\label{eta}
\eta = \frac{4 \pi r^{2} u}{L_{\odot}}
\sum_{j} n_{sw~j} m_{sw~j} c^{2} ~,
\end{equation}
here, $u$ is the speed of the solar wind with respect to the Sun,
$u$ $=$ 450 km/s, $m_{sw~j}$ and $n_{sw~j}$,
are the masses and concentrations of the solar wind particles of $j$-th
type at a distance $r$ from the Sun, respectively. $\eta$ $=$ 0.38 for
the Sun \citep{covsw}. $c_{0i}$ is the drag coefficient for
the dust particle at rest with respect to the Sun \citep{baines}
\begin{eqnarray}\label{cd0}
c_{0i} &=& \frac{1}{\sqrt{\pi}}
      \left ( \frac{1}{s_{0i}} + \frac{1}{2 s_{0i}^{3}} \right )
      \mbox{e}^{-s_{0i}^{2}}
\nonumber \\
& &   + ~\left ( 1 + \frac{1}{s_{0i}^{2}} - \frac{1}{4 s_{0i}^{4}} \right )
\mbox{erf}(s_{0i})
\nonumber \\
& &   + ~\left ( 1 - \delta_{i} \right )
      \left ( \frac{T_{d}}{T_{i}} \right )^{1/2}
      \frac{\sqrt{\pi}}{3s_{0i}} ~,
\end{eqnarray}
here, erf$(s_{i})$ is the error function $\mbox{erf}(s_{0i})$ $=$
$2 / \sqrt{\pi}$ $\int_{0}^{s_{0i}}$ $\mbox{e}^{-t^{2}}$ $dt$, $\delta_{i}$ is
the fraction of impinging particles specularly reflected at the surface
(for the resting particles, there is assumed diffuse reflection)
\citep{baines,gustafson}, $T_{d}$ is the temperature of the dust grain,
and $T_{i}$ is the temperature of the $i$-th gas component, $s_{0i}$
is the molecular speed ratio for the dust particle at rest with respect
to the Sun (Mach number)
\begin{equation}\label{s0}
s_{0i} = \sqrt{\frac{m_{i}}{2kT_{i}}} v_{F} ~,
\end{equation}
$m_{i}$ is the mass of the neutral atoms of type $i$,
$k$ is the Boltzmann's constant, $\vec{v}_{F}$ $=$ $(v_{FX},v_{FY},v_{FZ})$
is the velocity of the IGF in the frame
associated with the Sun, $v_{F}$ $=$ $\vert \vec{v}_{F} \vert$,
$\gamma_{i}$ is the collision parameter,
\begin{equation}\label{cp}
\gamma_{i} = n_{i} \frac{m_{i}}{m} A ~,
\end{equation}
here, $n_{i}$ is the concentration of the interstellar neutral
atoms of type $i$, and $A$ $=$ $\pi {R}^{2}$ is the geometrical cross
section of the dust particle with mass $m$, $p_{\beta}$ $=$
$a_{\beta} (1 - e_{\beta}^{2})$,
\begin{equation}\label{sigma}
\sigma_{\beta} = \frac{\sqrt{\mu ( 1 - \beta ) / p_{\beta}}}{v_{F}} ~,
\end{equation}
the parameter $g_{i}$ describes variability of the drag coefficient
\citep{dyncd}
\begin{eqnarray}\label{gexpanded}
g_{i} &=& \frac{1}{c_{0i}} \biggl [ \frac{1}{\sqrt{\pi}}
      \left ( \frac{1}{s_{0i}} - \frac{3}{2s_{0i}^{3}} \right )
      \mbox{e}^{-s_{0i}^{2}}
\nonumber \\
& &   + ~\left ( 1 - \frac{1}{s_{0i}^{2}} + \frac{3}{4s_{0i}^{4}} \right )
      \mbox{erf}(s_{0i}) \biggr ]
\end{eqnarray}
and
\begin{eqnarray}\label{SIC}
S_{\beta} &=& ( \cos \Omega_{\beta} \cos \omega_{\beta} -
      \sin \Omega_{\beta} \sin \omega_{\beta} \cos i_{\beta} ) v_{FX}
\nonumber \\
& &   + ~( \sin \Omega_{\beta} \cos \omega_{\beta} +
      \cos \Omega_{\beta} \sin \omega_{\beta} \cos i_{\beta} ) v_{FY}
\nonumber \\
& &   + ~\sin \omega_{\beta} \sin i_{\beta} v_{FZ} ~,
\nonumber \\
I_{\beta} &=& ( - \cos \Omega_{\beta} \sin \omega_{\beta} -
      \sin \Omega_{\beta} \cos \omega_{\beta} \cos i_{\beta} ) v_{FX}
\nonumber \\
& &   + ~( - \sin \Omega_{\beta} \sin \omega_{\beta} +
      \cos \Omega_{\beta} \cos \omega_{\beta} \cos i_{\beta} ) v_{FY}
\nonumber \\
& &   + ~\cos \omega_{\beta} \sin i_{\beta} v_{FZ} ~,
\nonumber \\
C_{\beta} &=& \sin \Omega_{\beta} \sin i_{\beta} v_{FX} -
      \cos \Omega_{\beta} \sin i_{\beta} v_{FY}
\nonumber \\
& &   + ~\cos i_{\beta} v_{FZ} ~.
\end{eqnarray}
If we use Eqs. (\ref{dadt_sys}), (\ref{dedt_sys}) and (\ref{didt_sys})
in Eq. (\ref{full}), we obtain
\begin{eqnarray}\label{evolutions}
\left \langle \frac{dC_{T}}{dt} \right \rangle &=&
      \frac{\partial C_{T}}{\partial e_{\beta}}
      \left \langle \frac{de_{\beta}}{dt} \right \rangle +
      \frac{\partial C_{T}}{\partial i_{\beta}}
      \left \langle \frac{di_{\beta}}{dt} \right \rangle
\nonumber \\
&=& \beta \frac{\mu}{c} \left ( 1 + \frac{\eta}{\bar{Q}'_{pr}} \right )
      \sqrt{\frac{1 - \beta}{a_{P}^{3} a^{3}}}
\nonumber \\
& &   \times ~\left [ \frac{( 1 - \beta )^{1/2} ( 2 + 3 e_{\beta}^{2} )}
      {2 ( 1 - e_{\beta}^{2} )^{3/2}}
      \left ( \frac{a_{P}}{a} \right )^{3/2} - \cos i_{\beta} \right ]
\nonumber \\
& &   + ~\sum_{i} c_{0i} \gamma_{i} v_{F}^{2}
      \sqrt{\frac{p_{\beta}}{\mu \left ( 1 - \beta \right )}}
\nonumber \\
& &   \times ~\Biggl \{
      \sqrt{\frac{\left ( 1 - \beta \right ) p_{\beta}}{a_{P}^{3}}}
      \Biggl [ \frac{3}{2} \frac{1}{v_{F}}
      \frac{e_{\beta}}{1 - e_{\beta}^{2}}
\nonumber \\
& &   \times ~\left ( C_{\beta} \cos \omega_{\beta} \sin i_{\beta} -
      I_{\beta} \cos i_{\beta} \right )
\nonumber \\
& &   - ~\sigma_{\beta} \cos i_{\beta} + \frac{1}{2}
      \frac{\sigma_{\beta} g_{i}}{v_{F}^{2}}
      \bigl ( C_{\beta} ( S_{\beta} \sin \omega_{\beta}
\nonumber \\
& &   + ~I_{\beta} \cos \omega_{\beta} ) \sin i_{\beta} -
      \left ( S_{\beta}^{2} + I_{\beta}^{2} \right )
      \cos i_{\beta} \bigr ) \Biggr ]
\nonumber \\
& &   + ~\frac{1 - \beta}{a_{\beta}} \sigma_{\beta}
      \Biggl [ 1 + \frac{g_{i}}{v_{F}^{2}}
      \frac{1 - \sqrt{1 - e_{\beta}^{2}}}{e_{\beta}^{2}}
\nonumber \\
& &   \times ~\left ( S_{\beta}^{2} +
      I_{\beta}^{2} \sqrt{1 - e_{\beta}^{2}} \right ) \Biggr ] \Biggr \} ~.
\end{eqnarray}
\begin{figure}
\begin{center}
\includegraphics[width=0.415\textwidth]{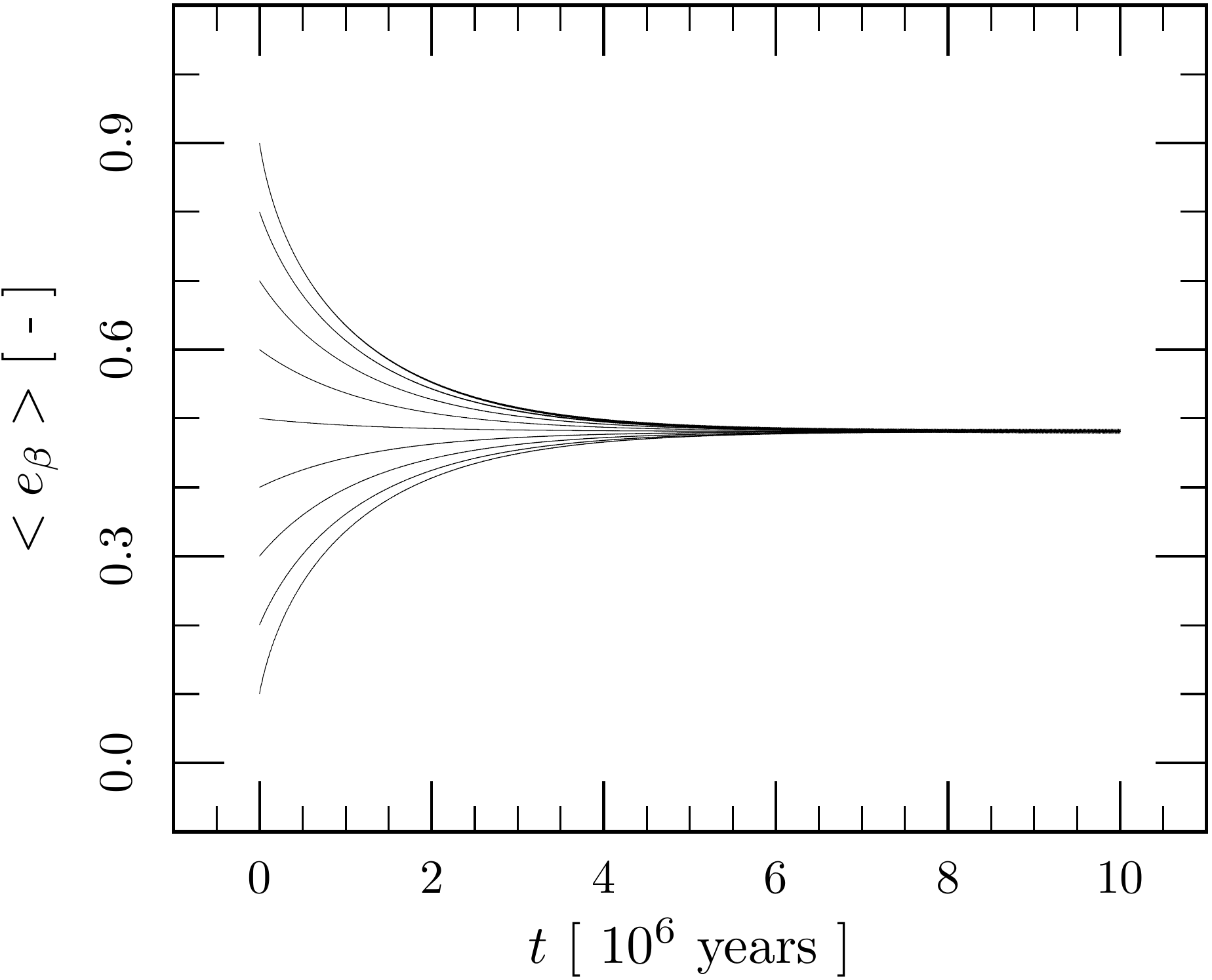}
\end{center}
\caption{Evolutions of eccentricity of a dust particle with
$R$ $=$ 2 $\mu$m, $\varrho$ $=$ 1 g/cm$^{3}$, and $\bar{Q}'_{pr}$ $=$ 1
captured in an exterior mean motion orbital 2/1 resonance with
Neptune under the action of the PR effect and radial solar wind
for various initial eccentricities.}
\label{F3}
\end{figure}
\begin{figure}
\begin{center}
\includegraphics[width=0.415\textwidth]{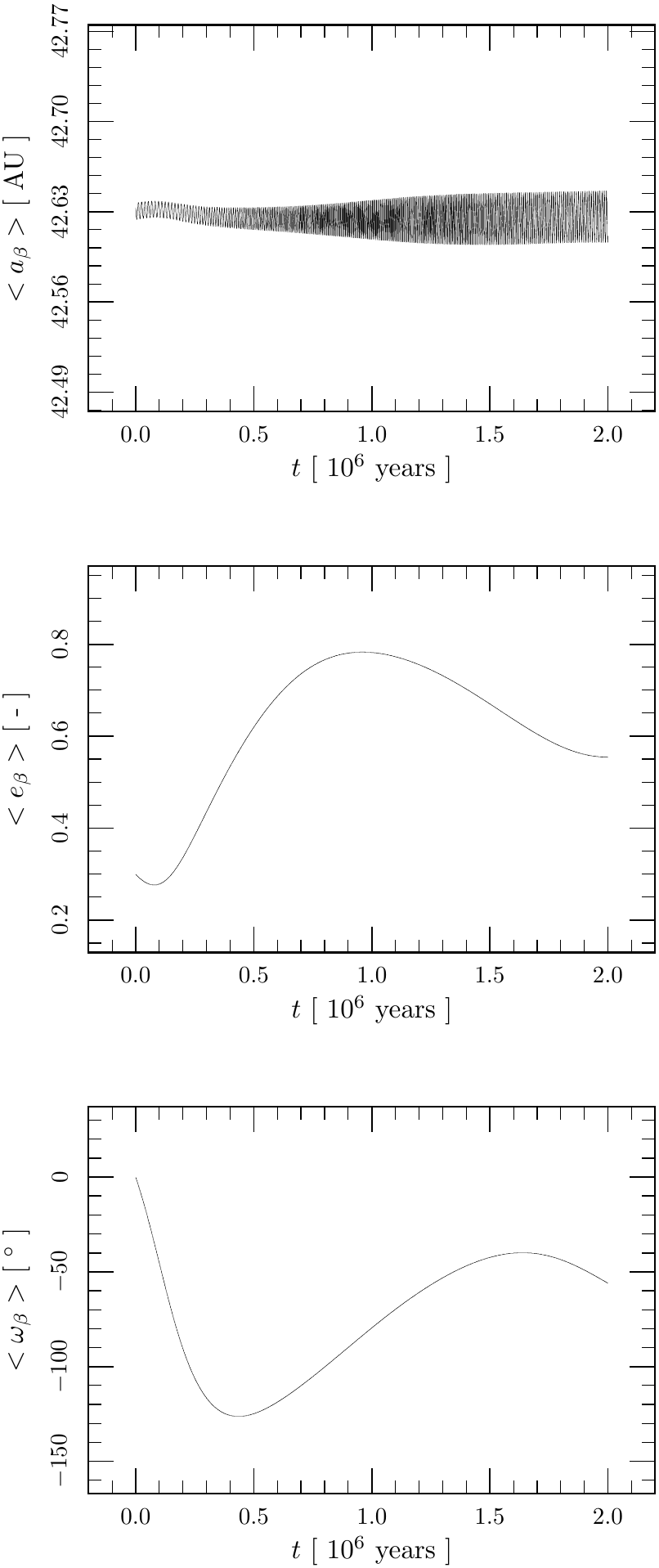}
\end{center}
\caption{Evolutions of the orbit averaged semi-major axis, eccentricity and
argument of perihelion of a dust particle with $R$ $=$ 2 $\mu$m,
$\varrho$ $=$ 1 g/cm$^{3}$, and $\bar{Q}'_{pr}$ $=$ 1
captured in an exterior mean motion orbital 2/1 resonance with
Neptune under the action of the PR effect, radial solar wind and
IGF in the planar case.}
\label{F4}
\end{figure}
This is the relation between $\langle de_{\beta} / dt \rangle$ and
$\langle di_{\beta} / dt \rangle$ which must hold during the orbital evolution
of the dust particle captured in an MMR. When the position vector and
the velocity vector of the dust particle lie in the planet orbital plane and
the interstellar gas velocity vector lies also in this plane, the dust
particle's motion is coplanar. In this case we have $i_{\beta}$ $=$ 0
and $\langle di_{\beta} / dt \rangle$ $=$ 0. Using these assumptions
we obtain for the secular time derivative of eccentricity from
Eq. (\ref{evolutions}) and Eq. (\ref{axes})
\begin{eqnarray}\label{planar}
\left \langle \frac{de_{\beta}}{dt} \right \rangle &=&
      \beta \frac{\mu}{c} \left ( 1 + \frac{\eta}{\bar{Q}'_{pr}} \right )
      \frac{( 1 - e_{\beta}^{2} )^{1/2}}{a_{\beta}^{2} e_{\beta}}
\nonumber \\
& &   \times ~\left [ 1 - \frac{2 + 3 e_{\beta}^{2}}
      {2 ( 1 - e_{\beta}^{2} )^{3/2}} \frac{n}{n_{P}} \right ]
\nonumber \\
& &   + ~\sum_{i} c_{0i} \gamma_{i} v_{F}^{2}
      \sqrt{\frac{p_{\beta}}{\mu \left ( 1 - \beta \right )}}
\nonumber \\
& &   \times ~\Biggl \{ \frac{3}{2} \frac{I_{\beta}}{v_{F}} +
      \frac{1 - e_{\beta}^{2}}{e_{\beta}} \sigma_{\beta}
      \Biggl [ 1 + \frac{1}{2} \frac{g_{i}}{v_{F}^{2}}
      \left ( S_{\beta}^{2} + I_{\beta}^{2} \right ) \Biggr ]
\nonumber \\
& &   - ~\frac{( 1 - e_{\beta}^{2} )^{1/2}}{e_{\beta}}
      \frac{n}{n_{P}} \sigma_{\beta}
      \Biggl [ 1 + \frac{g_{i}}{v_{F}^{2}}
      \frac{1 - \sqrt{1 - e_{\beta}^{2}}}{e_{\beta}^{2}}
\nonumber \\
& &   \times ~\left ( S_{\beta}^{2} +
      I_{\beta}^{2} \sqrt{1 - e_{\beta}^{2}} \right ) \Biggr ] \Biggr \} ~.
\end{eqnarray}

\section{Numerical results}
\label{sec:numeric}

In this section we want to use numerical solutions of equation of motion
in order to compare numerical results with the analytical results derived in
the previous section and to find some main properties of dust particle
orbital evolution in an MMR with a planet under the action of
the solar radiation and IGF.

\subsection{Equation of motion}
\label{sec:EOM}

If we consider the gravitation of the Sun, gravitation of one planet,
PR effect, radial solar wind and IGF, then equation of
motion of the dust particle has form \citep{covsw,Icarus,baines}
\begin{eqnarray}\label{EOM}
\frac{d \vec{v}}{dt} &=& - \frac{\mu}{r^{2}}
      \left ( 1 - \beta \right ) \vec{e}_{R}
\nonumber \\
& &   - ~\beta \frac{\mu}{r^{2}}
      \left ( 1 + \frac{\eta}{\bar{Q}'_{pr}} \right )
      \left ( \frac{\vec{v} \cdot \vec{e}_{R}}{c}
      \vec{e}_{R} + \frac{\vec{v}}{c} \right )
\nonumber \\
& &   - ~\sum_{i} c_{Di} \gamma_{i}
      \vert \vec{v} - \vec{v}_{F} \vert
      \left ( \vec{v} - \vec{v}_{F} \right )
\nonumber \\
& &   - ~\frac{G M_{P}} {\vert \vec{r} - \vec{r_{P}} \vert^{3}}
      (\vec{r} - \vec{r_{P}})
\nonumber \\
& &   - ~\frac{G M_{P}}{\vert \vec{r_{P}} \vert^{3}}
      \vec{r_{P}} ~,
\end{eqnarray}
where $\vec{v}$ $=$ $d \vec{r} / dt$ is the velocity of the dust particle,
$\vec{e}_{R}$ $=$ $\vec{r} / r$ is the radial unit vector,
$c_{Di}$ is the drag coefficient \citep{baines}
\begin{eqnarray}\label{cd}
c_{Di} &=& \frac{1}{\sqrt{\pi}}
      \left ( \frac{1}{s_{i}} + \frac{1}{2 s_{i}^{3}} \right )
      \mbox{e}^{-s_{i}^{2}}
\nonumber \\
& &   + ~\left ( 1 + \frac{1}{s_{i}^{2}} - \frac{1}{4 s_{i}^{4}} \right )
      \mbox{erf}(s_{i})
\nonumber \\
& &   + ~\left ( 1 - \delta_{i} \right )
      \left ( \frac{T_{d}}{T_{i}} \right )^{1/2}
      \frac{\sqrt{\pi}}{3s_{i}} ~,
\end{eqnarray}
$s_{i}$ is the molecular speed ratio
\begin{equation}\label{s}
s_{i} = \sqrt{\frac{m_{i}}{2kT_{i}}} \vert \vec{v} - \vec{v}_{F} \vert
\end{equation}
and $\vec{r_{P}}$ is the position vector of the planet with respect
to the Sun.
\begin{figure}
\begin{center}
\includegraphics[width=0.415\textwidth]{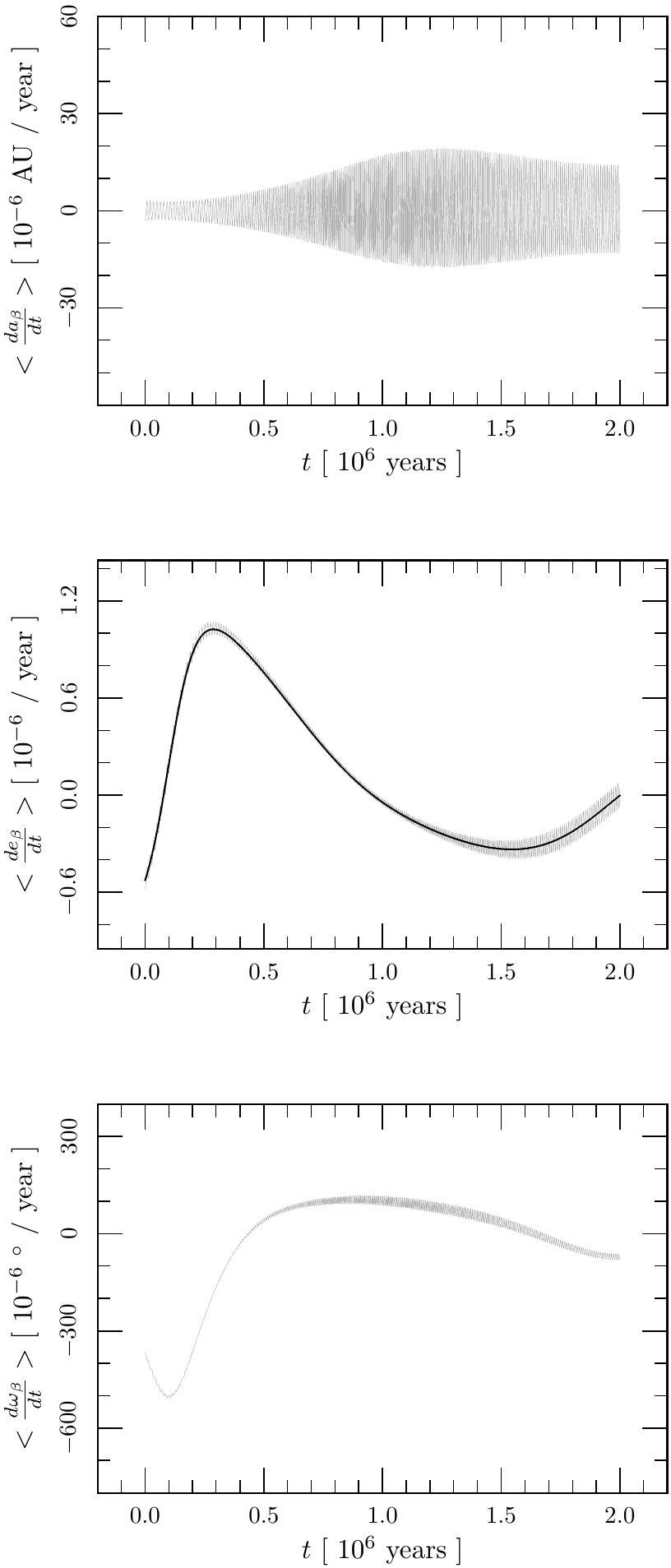}
\end{center}
\caption{Evolutions of the orbit averaged time derivatives of semi-major axis,
eccentricity and argument of perihelion during the numerical solution of
Eq. (\ref{EOM}) depicted in Fig. \ref{F4} (grey line). The evolution of
the orbit averaged time derivative of eccentricity is compared with
the secular time derivative of eccentricity obtained from Eq. (\ref{planar})
(black line).}
\label{F5}
\end{figure}

\subsection{Comparing analytical and numerical results in the three-dimensional
case}
\label{sec:3D}

From numerical solution of Eq. (\ref{EOM}) the orbit averaged time
derivative of Tisserand parameter can be obtained. We used the numerical
solution of Eq. (\ref{EOM}) and Eq. (\ref{evolutions}) for comparison
of the numerical and analytical values of the secular time derivatives
of Tisserand parameter. In the IGF we
considered the primary and secondary populations
of neutral hydrogen atoms and neutral helium atoms. The primary population
of neutral hydrogen atoms and neutral helium atoms represent
the original atoms of the IGF which
penetrate into the heliosphere. The secondary population of neutral
hydrogen atoms comprises the former protons from the IGF
that acquired electrons from interstellar H$^{\circ}$ between the bow shock
and the heliopause \citep{frisch,alouani}. We adopted the following
parameters for these components in the IGF.
$n_{1}$ $=$ 0.059 cm$^{-3}$ and $T_{1}$ $=$ 6100 K
for the primary population of neutral hydrogen \citep{frisch},
$n_{2}$ $=$ 0.059 cm$^{-3}$ and $T_{2}$ $=$ 16500 K
for the secondary population of neutral hydrogen \citep{frisch} and finally
$n_{3}$ $=$ 0.015 cm$^{-3}$ and $T_{3}$ $=$ 6300 K for the neutral helium
\citep{lallement}. We have assumed that the interstellar
gas velocity vector is equal for all components and identical to
the velocity vector of the neutral helium entering the Solar system.
The neutral helium enter the Solar system with a
speed of about $v_{F}$ $=$ 26.3 km s$^{-1}$ \citep{lallement}, and
arrive from the direction of $\lambda_{ecl}$ $=$ 254.7$^{\circ}$
(heliocentric ecliptic longitude) and $\beta_{ecl}$ $=$ 5.2$^{\circ}$
(heliocentric ecliptic latitude; \citealt{lallement}).
The components of the velocity in the ecliptic coordinates with
the $x$-axis aligned towards the actual equinox $\vec{v}_{F}$ $=$
$-$ 26.3 km/s [$\cos(254.7^{\circ}) \cos(5.2^{\circ})$,
$\sin(254.7^{\circ}) \cos(5.2^{\circ})$, $\sin(5.2^{\circ})$] were
transformed into a coordinate system with the $xy$-plane lying
in the orbital plane of planet Neptune and the $x$-axis lying in the ecliptic
plane. The orbital elements of the dust particle were determined
in this reference coordinate system. The drag coefficients for Eq. (\ref{EOM})
were calculated from Eq. (\ref{cd}). We assumed that the atoms are specularly
reflected at the surface of the dust grain ($\delta_{i}$ $=$ 1).
The planet was initially located on the $x$-axis. Initial semimajor axis
of the dust particle is, in general, computed from
the relation $a_{\beta~in}$ $=$ $a_{\beta ~n_{P} / n}$ $+$ $\triangle$,
where $a_{\beta ~n_{P} / n}$ is defined by Eq. (\ref{axes}) (with $\beta$
calculated from Eq. \ref{beta}) and $\triangle$ is a shift from
the exact resonant semimajor axis. As the initial conditions for a dust
particle with $R$ $=$ 2 $\mu$m, mass density $\varrho$ $=$ 1 g/cm$^{3}$,
and $\bar{Q}'_{pr}$ $=$ 1, we used $a_{\beta~in}$ $=$ $a_{\beta ~2 / 1}$
$+$ 0.001 AU, $e_{\beta~in}$ $=$ 0.5, $\omega_{\beta~in}$
$=$ 60$^{\circ}$, $\Omega_{\beta~in}$ $=$ 40$^{\circ}$, and $i_{\beta~in}$
$=$ 20$^{\circ}$. The initial true anomaly of the dust
particle was $f_{\beta~in}$ $=$ 345$^{\circ}$.
From the numerical solution of Eq. (\ref{EOM}) we obtained evolutions
of the orbit averaged orbital elements and the orbit averaged time derivatives
of orbital elements. Values of Tisserand parameter were determined in every
time step from the oscular orbital elements. From these values evolution
of the orbit averaged Tisserand parameter and the orbit averaged time
derivative of Tisserand parameter was also obtained. Evolutions obtained from
the numerical solution of Eq. (\ref{EOM}) are depicted in Fig. \ref{F1}
(black line) and Fig. \ref{F2} (grey line). We must note that the evolution
duration 10$^6$ years requires the size of an interstellar gas
cloud 26.9 pc in the direction of the interstellar gas velocity vector
(constant velocity with magnitude 26.3 km/s is assumed) and such
situation cannot always occur in the real galactic environment.
To compare the numerical results with our analytical theory we calculated
values of the secular time derivative of Tisserand parameter from
Eq. (\ref{evolutions}) using the numerically calculated secular orbital
elements. The obtained evolution is depicted in the bottom right panel
of Fig. \ref{F2} using black line. The lines are practically overlapping
each other. We also calculated values of the secular time derivative
of eccentricity from equation (see Eq. \ref{evolutions})
\begin{equation}\label{comparison}
\left \langle \frac{de_{\beta}}{dt} \right \rangle =
\frac{1}{\partial C_{T} / \partial e_{\beta}}
\left \langle \frac{dC_{T}}{dt} \right \rangle -
\frac{\partial C_{T} / \partial i_{\beta}}
{\partial C_{T} / \partial e_{\beta}}
\left \langle \frac{di_{\beta}}{dt} \right \rangle ~,
\end{equation}
where $\left \langle dC_{T} / dt \right \rangle$ was calculated from
Eq. (\ref{evolutions}) and as $\left \langle di_{\beta} / dt \right \rangle$
we used the values obtained from numerical solution of Eq. (\ref{EOM})
and depicted in the bottom left panel of Fig. \ref{F2}. Obtained evolution
is shown in the top right panel of Fig. \ref{F2} with black line.
The analytical and numerical results are in excellent agreement.

\subsection{Comparing analytical and numerical results in the planar case}
\label{sec:2D}

In the planar case it is possible to compare directly the secular time
derivative of eccentricity obtained from numerical solution with the same
result obtained from analytical theory.

In the case when the dust particle captured in an MMR is only under
the action of the PR effect and radial solar wind, secular time
derivative of eccentricity is given by the first term in
Eq. (\ref{planar}), namely
\begin{eqnarray}\label{zook}
\left \langle \frac{de_{\beta}}{dt} \right \rangle &=&
      \beta \frac{\mu}{c} \left ( 1 + \frac{\eta}{\bar{Q}'_{pr}} \right )
      \frac{( 1 - e_{\beta}^{2} )^{1/2}}{a_{\beta}^{2} e_{\beta}}
\nonumber \\
& &   \times ~\left [ 1 - \frac{2 + 3 e_{\beta}^{2}}
      {2 ( 1 - e_{\beta}^{2} )^{3/2}} \frac{n}{n_{P}} \right ] ~.
\end{eqnarray}
Function $l(e)$ $=$ $( 2 + 3 e_{\beta}^{2} )$ $/$
$[ 2 ( 1 - e_{\beta}^{2} )^{3/2} ]$ is an increasing function of
eccentricity. Function obtains values from $l(0)$ $=$ 1 to
$\lim_{e \to 1} l(e)$ $=$ $\infty$. Therefore, the secular eccentricity is
always a decreasing function in an interior resonance. In an exterior
resonance such value of eccentricity $e_{\beta c}$ exists that
\begin{equation}\label{critical}
l(e_{\beta c}) = \frac{2 + 3 e_{\beta c}^{2}}
{2 ( 1 - e_{\beta c}^{2} )^{3/2}} = \frac{n_{P}}{n} = \frac{p + q}{p} ~.
\end{equation}
For $e_{\beta}$ $<$ $e_{\beta c}$ the secular eccentricity asymptotically
increases to $e_{\beta c}$ and for $e_{\beta}$ $>$ $e_{\beta c}$ the secular
eccentricity asymptotically decreases to $e_{\beta c}$. In the special
case mean motion 1/1 resonance secular eccentricity asymptotically decreases
to $e_{\beta c}$ $=$ 0 \citep{AA2,AA3}. To visualise the secular evolution
of eccentricity of the dust particle captured in an MMR only under the action
of the PR effect and radial solar wind we numerically solved Eq. (\ref{EOM})
without the third term. The planet was initially on the $x$-axis.
We used dust particle with $R$ $=$ 2 $\mu$m,
$\varrho$ $=$ 1 g/cm$^{3}$ and $\bar{Q}'_{pr}$ $=$ 1. The initial
semimajor axis was $a_{\beta~in}$ $=$ $a_{\beta ~2 / 1}$ $+$ 0.001 AU.
The initial eccentricities and arguments of perihelion were
0.1, 0.2, 0.3, 0.4, 0.5, 0.6, 0.7, 0.8, 0.9
and
260$^{\circ}$, 275$^{\circ}$, 285$^{\circ}$,
295$^{\circ}$, 295$^{\circ}$, 290$^{\circ}$,
280$^{\circ}$, 265$^{\circ}$, 250$^{\circ}$,
respectively. The initial true anomaly was $f_{\beta~in}$ $=$ 0
for all particles. Results are depicted in Fig. \ref{F3}.
The limiting value $e_{\beta c}$ for mean motion 2/1 resonance
obtained from Eq. (\ref{critical}) is $e_{\beta c}$ $\approx$ 0.4812.
As can be easily seen from Fig. \ref{F3} the secular eccentricity
really approaches this value.

An orbital evolution of the dust particle under the action of
the PR effect, radial solar wind and IGF in the planar case is
depicted in Figs. \ref{F4} and \ref{F5}.
We used the same parameters for various gas components of the IGF,
initial position of the planet and properties of the dust grain as in
Figs. \ref{F1} and \ref{F2}. The planar case was attained by a rotation
of the interstellar gas velocity vector into the orbital plane of Neptune
around axis lying in the orbital plane of Neptune and perpendicular
to the interstellar gas velocity vector. The rotation angle was
3.7$^{\circ}$. Initial orbital parameters of the dust
grain were $a_{\beta~in}$ $=$ $a_{\beta ~2 / 1}$ $+$ 0.001 AU,
$e_{\beta~in}$ $=$ 0.3, $\omega_{\beta~in}$ $=$ 0 and
$f_{\beta~in}$ $=$ 160$^{\circ}$. As can be seen in Fig. \ref{F4}
in this case the secular evolution of eccentricity is not a monotonic
function of time. The secular time derivatives of eccentricity obtained
from this numerical integration and from Eq. (\ref{planar}) are
compared in Fig. \ref{F5}. Again analytical theory and numerical
solution are in excellent agreement.
\begin{figure*}
\begin{center}
\includegraphics[width=0.9\textwidth]{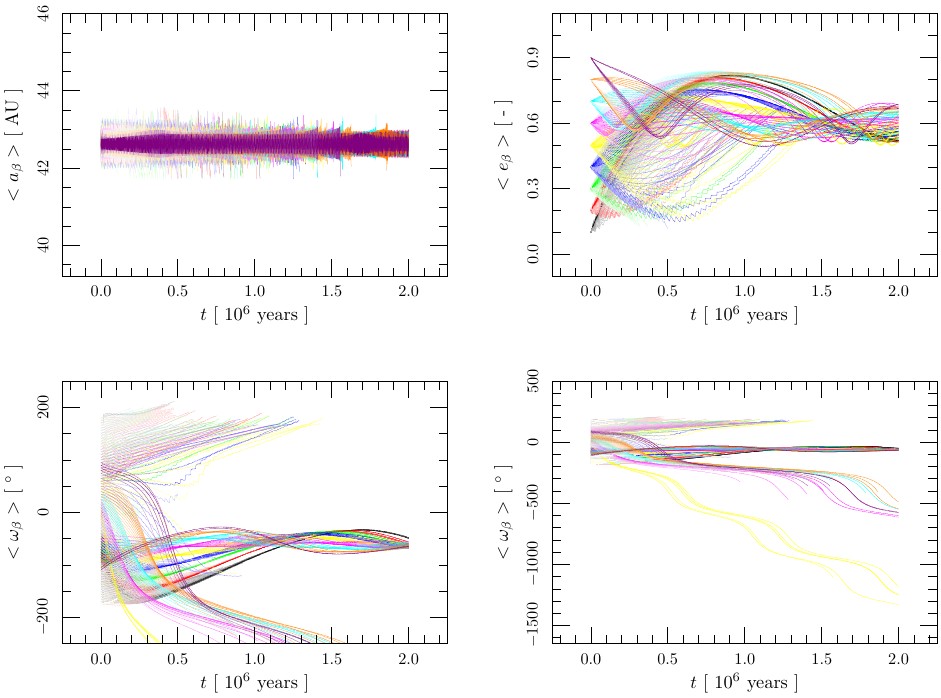}
\end{center}
\caption{Evolutions of the orbit averaged Keplerian orbital elements of a dust
particle with $R$ $=$ 2 $\mu$m, $\varrho$ $=$ 1 g/cm$^{3}$, and
$\bar{Q}'_{pr}$ $=$ 1 captured in an exterior mean motion orbital 2/1
resonance with Neptune under the action of the PR effect, radial solar
wind and IGF in the planar case. The initial values of
orbital parameters were $a_{\beta~in}$ $=$ $a_{\beta ~2 / 1}$ $+$ 0.001 AU,
$e_{\beta~in}$ $\in$ \{0.1, 0.2, ..., 0.9\}, $\omega_{\beta~in}$ $\in$
\{0, 5$^{\circ}$, 10$^{\circ}$, ..., 355$^{\circ}$\} and
$f_{\beta~in}$ $=$ 0. Depicted are only the evolutions with the capture time
longer than 10$^{5}$ years (see Fig. \ref{F7}).}
\label{F6}
\end{figure*}
\begin{figure*}
\begin{center}
\includegraphics[width=0.9\textwidth]{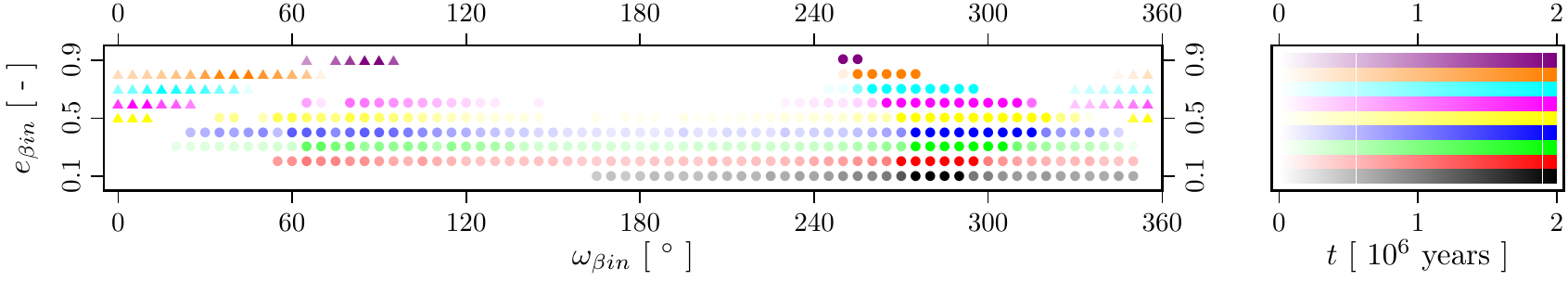}
\end{center}
\caption{The initial eccentricities and arguments of perihelion
of the evolutions depicted in Fig. \ref{F6}. Colour of each marker,
on a position given by the initial eccentricity and argument of perihelion,
corresponds the capture time of obtained evolution. The capture times are
from 0 to 2 $\times$ 10$^{6}$ years and the colours are from white to
the darkest shade of a given colour, respectively. If the capture time is
shorter than 10$^{5}$ years, then the corresponding position is empty.
The initial conditions of evolutions belonging to the second group
(evolutions with the fast monotonic shift of perihelion and oscillations
of eccentricity) are marked with triangles (see text).}
\label{F7}
\end{figure*}

\subsection{Qualitative properties of orbital evolution in the planar case}
\label{sec:quali}

In order to find qualitative properties of the dust particle orbital
evolution in MMRs under the action of the PR effect, radial solar
wind and IGF numerical solution Eq. (\ref{EOM})
for many various initial conditions is needed. Calculation of
the variable drag coefficients according to Eq. (\ref{cd}) is very time
consuming therefore we used an approximation that the drag coefficients are
constant. If the inequality $\vert \vec{v} \vert$ $\ll$ $v_{F}$ holds during
orbit, then this approximation is usable (see \citealt{dyncd}). Many sets
of initial conditions in Eq. (\ref{EOM}) with the constant drag
coefficients were numerically solved in order to study evolution
of the dust particle in an exterior MMR. One such set of numerical
solutions is depicted in Fig. \ref{F6}. We used the same properties
of the IGF, initial position of the planet and
properties of the dust grain as in Figs. \ref{F4}
and \ref{F5}. Initial orbital parameters of the dust grain were
$a_{\beta~in}$ $=$ $a_{\beta ~2 / 1}$ $+$ 0.001 AU,
$e_{\beta~in}$ $\in$ \{0.1, 0.2, ..., 0.9\}, $\omega_{\beta~in}$ $\in$
\{0, 5$^{\circ}$, 10$^{\circ}$, ..., 355$^{\circ}$\} and
$f_{\beta~in}$ $=$ 0. Therefore, we obtained 648 individual evolutions.
Not all 648 evolutions are depicted in Fig. \ref{F6}. An evolution is
depicted in Fig. \ref{F6} only if the dust particle was captured
and remained in the resonance longer than 10$^{5}$ years.
Each evolution depicted in Fig. \ref{F6} is represented by a marker
in Fig. \ref{F7}. Position of each marker corresponds to the initial
eccentricity and initial argument of perihelion. Colour of each marker
corresponds to the capture time for given evolution. White colour would
theoretically correspond to zero capture time and the darkest shade
of a given colour corresponds to the capture time 2 $\times$ 10$^{6}$ years.
A place in Fig. \ref{F7} without marker corresponds to the initial
eccentricity and initial argument of perihelion for which
the dust particle was not captured in the resonance or remained in
the resonance shorter than 10$^{5}$ years. 393 evolutions is depicted
in Fig. \ref{F6}. From many such sets of numerical solutions as set
depicted in Fig. \ref{F6} we found that two main groups of orbital
evolution in an exterior MMR under the action of the PR effect, radial
solar wind and IGF exist. As was already mentioned all evolutions depicted
in Fig. \ref{F6} have the initial true anomaly equal to zero.
For sets with different initial value of the true anomaly are these two
main groups preserved. These two groups have also longest capture times.

Orbital evolutions in the first group are characterised
by approach of the eccentricity and argument of
perihelion to some ``constant'' values. We found that these values
are not exactly constant but in comparison with the evolution before
this ``stabilisation'' are relatively slowly changing. The stabilised
direction of the position vector of the orbit's perihelion is almost parallel
with direction of the IGF velocity vector in Fig. \ref{F6}.
The parallel case corresponds to the argument of perihelion $\omega_{\beta}$
$\approx$ -57.16$^{\circ}$ in Fig. \ref{F6}. Many evolutions approach to
this value of $\omega_{\beta}$ as can be seen in the bottom left panel
of Fig. \ref{F6}. $I_{\beta}$ $=$ 0 in the parallel case (see Eq. \ref{SIC}).
If we use in Eqs. (\ref{dadt_sys})-(\ref{didt_sys}) condition
$\sigma_{\beta}$ $=$ 0, we obtain the secular time derivatives of
orbital elements caused by the PR effect, radial solar wind and
IGF described by a constant acceleration.
If the dust particle is only under the action of the solar gravity
(which can be reduced using the radial Keplerian term of the PR effect) and
the constant acceleration caused by the IGF, then the secular
orbital motion can be completely solved analytically (see \citealt{fgf,bera}).
Perhaps, the following idea can appear: for constant acceleration caused
by the IGF ($\sigma_{\beta}$ $=$ 0) the stabilisation
of eccentricity would correspond to trivial solution of Eq. (\ref{planar})
$I_{\beta}$ $=$ 0 and $e_{\beta}$ $=$ $e_{\beta c}$ (Eq. \ref{critical}).
Therefore, we solved Eq. (\ref{EOM}) with acceleration caused by
the IGF described by a constant vector (dependence
on particle velocity in the acceleration was not included) for
a set of initial conditions identical to the set used in Fig. \ref{F6}.
The stabilisation occurred also in this case. However, the argument of
perihelion into which perihelia approached was not equal to -57.16$^{\circ}$
but -63$^{\circ}$. Another important difference was in the evolution
eccentricity. The eccentricities approached to $e_{\beta}$ $\approx$ 0.61 and
not to $e_{\beta c}$ $\approx$ 0.4812. Therefore, at least in this simplified
case, the stabilisation does not correspond to trivial solution
of Eq. (\ref{planar}) $I_{\beta}$ $=$ 0 and $e_{\beta}$ $=$ $e_{\beta c}$.

Orbital evolutions in the second group are characterised by a fast monotonic
shift of perihelion and oscillations of eccentricity. Typical examples
of this group are depicted in the bottom right panel of Fig. \ref{F6} as
rapidly decreasing curves. The initial conditions which lead to this
behaviour are marked with triangles in Fig. \ref{F7}.

\section{Conclusion}
\label{sec:conclusion}

We investigated the orbital evolution of a dust particle captured in an MMR
with a planet in a circular orbit under the action of the PR effect, radial
stellar wind and IGF. The secular time derivative of Tisserand
parameter is analytically derived for arbitrary orbit orientation using
previously derived secular time derivatives of Keplerian orbital elements
caused by the PR effect, radial stellar wind and IGF.
From the secular time derivative of Tisserand parameter a relation between
the secular time derivatives of eccentricity and inclination can be obtained.
In the planar case we derived directly the secular time derivative
of eccentricity.

We numerically solved equation of motion of the dust particle in order to
compare the analytically derived results with the numerically obtained results
in the three-dimensional case and also in the planar case. In both cases
analytical and numerical results are in excellent agreement.
This implies that the theory developed in \citet{LZ1997} for
the PR effect and radial solar wind can be generalised for arbitrary
non-gravitational effects with known secular time derivatives.

If the dust particle is captured in an exterior MMR in
the planar CR3BP with the PR effect and radial stellar
wind, then the secular eccentricity is a monotonic function of time.
If we take into account also an IGF, complicated behaviour
with oscillations shows. This is most likely caused by dependence of
the secular time derivative of eccentricity on the argument of pericentre
due to directional character of the acceleration caused by the IGF.
However, qualitative properties of the secular evolution
of eccentricity and argument of pericentre can be determined. We found that
two main groups exist. In the first group the eccentricity and argument
of pericentre approach to some values. In the second group the eccentricity
oscillates and argument of pericentre rapidly shifts.

\label{lastpage}

\end{document}